\documentclass[runningheads]{llncs}
\usepackage{graphicx}
\usepackage{url}
\usepackage{color}

\usepackage{comment} 
\usepackage{subfigure}
\usepackage{booktabs}

\input{psfig.sty}

\begin{document}

\pagestyle{headings}

\mainmatter

\title{ Interplay of Game Incentives, Player Profiles and Task Difficulty in Games with a Purpose }
\author{ Gloria {Re Calegari} \and Irene Celino }
\institute{Cefriel -- Politecnico of Milano, Viale Sarca 226, 20126 Milano, Italy
 \\ \email{\scriptsize{\{gloria.re,irene.celino\}@cefriel.com} }
}

\titlerunning{ Interplay of Game Incentives, Player Profiles and Task Difficulty in GWAPs }
\authorrunning{ Gloria {Re Calegari} \and Irene Celino }

\maketitle

\begin{abstract}
How to take multiple factors into account when evaluating a Game with a Purpose? 
How is player behaviour or participation influenced by different incentives? 
How does player engagement impact  their accuracy in solving tasks? 
In this paper, we present a detailed investigation of multiple factors affecting the evaluation of a GWAP and we show how they impact on the achieved results. We inform our study with the experimental assessment of a GWAP designed to solve a multinomial classification task.
\end{abstract}

\section{Introduction}

Games with a Purpose~\cite{von2008designing} are a well-known Human Computation approach~\cite{law2011human} to encourage users to execute tasks with an entertaining reward. While several metrics are proposed in literature to evaluate the ability of GWAPs to achieve their intended purpose, there is a large number of factors that influences their success and effectiveness. 

In order to fully understand the strengths as well as the weaknesses of a GWAP, we propose an approach that takes into account \emph{player characteristics} (reliability, participation, behaviour and accuracy), \emph{game aspects} (playing incentive, playing style and game nature) and \emph{features of the task} to be solved (level of difficulty and variety).
Our goal is to investigate the interplay between those different factors, by proposing a multi-faceted analysis framework that allows for a deep assessment and understanding of the efficacy of a GWAP to achieve its purpose. We apply the proposed framework to a specific GWAP to show the empirical results and the insights that can be drawn through our approach.

The original contributions of this paper are: (1) an extension of traditional GWAP metrics to take temporal evolution and incentive effects into account; (2) a comparison of engagement metrics and engagement profiles with non-gaming citizen science; and (3) the definition of GWAP-specific engagement profiles and their interplay with different factors (incentive, task difficulty and task variety).

The remainder of the paper is organized as follows: Section~\ref{sec:related} illustrates the main related work; Section~\ref{sec:NK} gives details about the GWAP that we use to exemplify our approach; in the following sections, we propose different evaluation methods, by extending state-of-the-art metrics: global GWAP metrics and interplay with incentive are adopted in Section~\ref{sec:analysis1}, Section~\ref{sec:engagement} offers a comparison with citizen science user engagement profiles and Section~\ref{sec:analysis2} proposes new GWAP player profiling driven by measures of participation and accuracy; finally, Section~\ref{sec:concl} concludes the paper.

\section{Related work}\label{sec:related}

The basic metrics to evaluate GWAPs~\cite{von2008designing,law2011human,von2004labeling} are global indicators computed as means over the entire data; while effective in summarizing the behaviour of GWAP players, those are very simple measures that do not tell the entire story: an analysis of data distribution and temporal evolution is usually required to get a deeper understanding of a GWAP. 

Some work exists on cross-feature analysis of GWAPs~\cite{singh2017lessons} and similarly on citizen science~\cite{sauermann2015crowd} and crowdsourcing~\cite{yang2016modeling}; our goal is to contribute to making such evaluation easier to replicate and reproduce. 

Participation incentives are usually classified as intrinsic or extrinsic motivation~\cite{ryan2000intrinsic}. Some comparative analysis of incentives exists for GWAPs~\cite{prestopnik2017gamers}, especially in contrast to different methods like micro-working~\cite{thaler2012experiment,feyisetan2015improving,feyisetan2017social} or machine learning~\cite{re2018human}. 
The effect of competition and tangible rewards on participation and quality of results has also been explored, both in the context of GWAPs~\cite{siu2014collaboration} and online citizen science campaigns~\cite{reeves2018game}, revealing the pros and cons of designing different motivation mechanisms.	

Other metrics to evaluate GWAPs can be borrowed from studies of social community~\cite{reeves2017crowd} and citizen science evolution~\cite{celino2017citizen};  in those cases, however, user participation's ``success'' is measured through simple indicators like number of participants and contributions, while a deeper investigation is needed to assess the effectiveness of participation. 
Behavioural studies in HCI research have investigated volunteer characterization in citizen science, defining engagement metrics and profiles~\cite{ponciano2015finding,aristeidou2017profiles}, which may or may not apply to GWAP players.

In the context of (paid) crowdsourcing, assessment is usually conducted in relation to  micro-work platforms~\cite{allahbakhsh2013quality}, in which important features are related to cost minimization~\cite{karger2014budget,han2017budgeted} which is out of scope with respect to our work. 

While Games with a Purpose are a well-known and widely adopted human computation method to involve users in task solution, a comprehensive assessment of their ability to address their ``purpose'' needs to take into account multiple factors affecting the game and the players. We therefore propose a multi-faceted analysis framework for GWAPs that includes game aspects, player characteristics and task features, with specific focus on the effect of game incentives on the overall GWAP efficacy.

\section{Use Case: the Night Knights GWAP}\label{sec:NK}

The GWAP that we will use as running example is Night Knights, an online game for the multinomial classification of images\footnote{Cf. \url{https://www.nightknights.eu/}. }. Pictures come from a massive public-domain dataset provided by NASA and they can be classified according to six different categories depending on their visual content. The classified images -- in particular those labeled with three of the six categories -- are then used in a subsequent scientific workflow in the field of astronomy and environmental sciences to measure light pollution effects (cf.~\cite{re2018human}).

The GWAP is inspired by the ESP game~\cite{von2004labeling}, because users play in random pairs according to an output-agreement mechanism~\cite{von2008designing}. The game adopts a repeated labeling approach~\cite{sheng2008get} by asking different players to classify the same image; conversely, the same image is never given twice to the same player. Night Knights is built on top of our open source software framework for GWAPs~\cite{re2018framework}.

\begin{figure}[bt]
	\hfill
	\subfigure[Classify an image]{\label{fig:game}\includegraphics[height=2.8in]{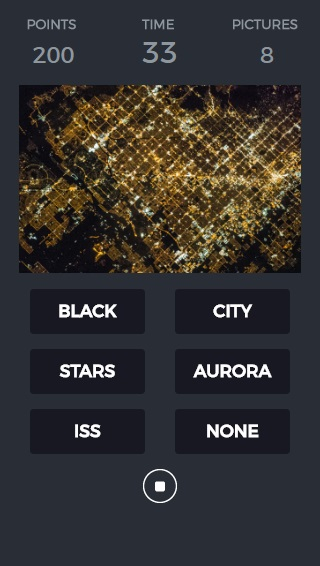}}
	\hfill
	\subfigure[Agreement]{\label{fig:agreement}\includegraphics[height=2.8in]{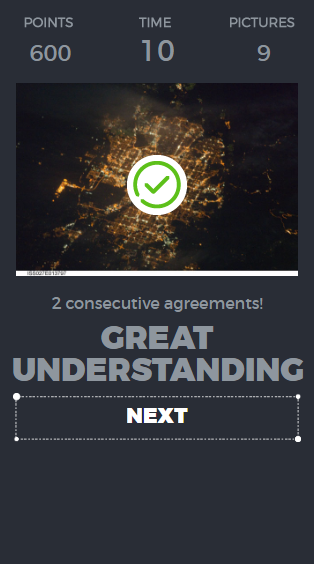}}
	\hfill
	\subfigure[Disagreement]{\label{fig:disagreement}\includegraphics[height=2.8in]{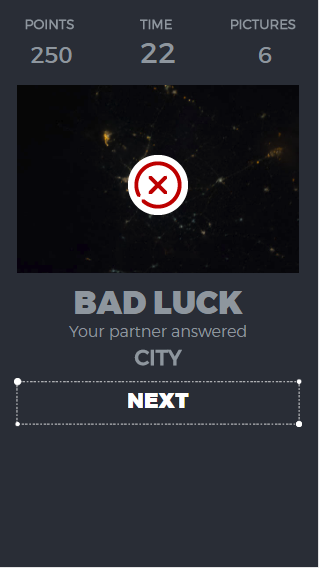}}
	\caption{Night Knights: the gameplay}
	\label{fig:game-play}
\end{figure}

The players visualize a picture and six buttons reporting the six possible categories (cf. Figure~\ref{fig:game-play}); the labeling task is therefore executed by clicking on the category that better fits the picture content. Each game round lasts  one minute, during which players can classify as many images as they can (as detailed in the following, on average 15 pictures are played per round); each time the two players agree, they gain points and level up in the game leaderboard; some badges are also assigned in special conditions as additional game intrinsic incentives.

Players' contributions are aggregated through an incremental truth inference algorithm~\cite{celino2018incremental} that (1) processes inputs as soon as they are provided, (2) weights players' answer with a round-specific reliability measure~\cite{celino2012linking} taking into account players' answers on control tasks (for which the ``true'' solution is known), and (3) dynamically adjusts the number of required contributions. Our truth inference approach accounts for the very nature of GWAPs, in which usually there is no ``deadline'' for contributing, players' varying attention can impact answer quality and task difficulty needs a dynamic estimation of the required number of repeated labeling.

In this paper, we use the data collected through Night Knights. The game was released in February 2017 and then it was more extensively advertised for a related competition whose winner joined the 2017 Summer Expedition to observe the Solar Eclipse in USA. 
The competition lasted about one month, from mid June to mid July 2017, and was addressed to all EU University students. After the end of the competition, the game has still been available online, but without any additional advertising. Overall, the data we analyse  was collected in 9 months, one month of competition and 4 months before and after it\footnote{Data is available with a CC-BY license at \url{http://ckan.stars4all.eu/}.}. 

In the following experimental sections, we apply a set of assessment methods on this game data. On the one hand, we exemplify the analyses we propose for a thorough multi-faceted assessment of GWAPs; on the other hand, we provide  concrete results from the evaluation of Night Knights, which are -- at least partially -- typical of GWAPs.

\section{Extending GWAP metrics}\label{sec:analysis1}

The main metrics adopted in literature~\cite{law2011human} to evaluate GWAPs are: 
\textbf{throughput}, computed as the average number of solved task per unit of time, 
\textbf{average life play} or ALP, i.e. the average time spent by each user playing the game, 
and \textbf{expected contribution} or EC, measured as average number of tasks solved by each player. 
A task is solved when player contributions, aggregated by the truth inference algorithm~\cite{zheng2017truth}, output a ``true'' solution.
Those indicators are global measures, as they are computed as mean values over the entire GWAP use. Hereafter, we extend this analysis by assessing the \emph{influence of different game incentives} and the \emph{evolution over time} of game-play and engagement. 

In particular, we investigate how player participation and GWAP results change with and without an extrinsic motivation such as a tangible reward~\cite{ryan2000intrinsic}. We analyse incentive effect in terms of both general statistics and specific metrics adopted in GWAP evaluation. 
We show that users participation can be highly influenced by the presence of an extrinsic motivation. 

\subsection{[Q1] How do user participation and GWAP results change with different incentives?}

In 9 months, Night Knights managed to engage about 650 users that played a substantial amount of time and classified almost 28,000 photos (cf. Table~\ref{table:gwap-metrics}). 

\begin{table}[t] 
	\centering
	\begin{tabular}{@{}lccc@{}}
		\toprule
		                                 & \textbf{~Before~} & \textbf{~During~} & \textbf{~After~} \\ 
		\midrule
		\textbf{time span (months)}      & 4               & 1               & 4              \\
		\textbf{classified images}       & 1,830           & 24,600          & 1,300          \\
		\textbf{contributions}           & 13,000          & 187,600         & 3,600          \\
		\textbf{users}                   & 285             & 174             & 174            \\
		\textbf{total play time (hours)} & 65              & 471             & 29             \\ 
		\textbf{throughput (tasks/hour)} & 69              & 212             & 113            \\
		\textbf{ALP (mins/user)}         & 5.5             & 65              & 4              \\
		\textbf{EC (tasks/user)}         & 6.4             & 141             & 7.5            \\		
		\bottomrule \\
	\end{tabular}
	\caption{Experimental results in the three periods (before, during and after the introduction of the extrinsic motivation)}
	\label{table:gwap-metrics}
	\vspace{-.5cm}
\end{table}

\begin{figure}[h]
	\vspace{-.5cm}
	\centering
	\includegraphics[width=.8\columnwidth]{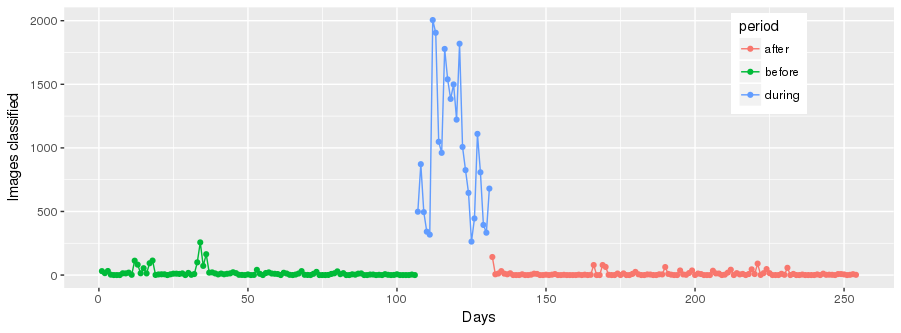}
	\caption{Number of images classified per day in the three periods}
	\label{fig:classified-per-day}
	\vspace{-.8cm}
\end{figure}

Measuring the main metrics in the three periods (\textit{before}, \textit{during} and \textit{after} the competition), we notice a significant increase of player participation during the competition, both in terms of given contributions and classified images (one order of magnitude higher with the additional incentive in both cases). This difference is clearly highlighted in Figure~\ref{fig:classified-per-day}, which shows the temporal evolution of the number of images classified per day.  The difference between throughput, ALP and EC in the competition and non-competition periods is statistically significant (t-test or Wilcoxon rank sum test at the 0.01 significance level). 
Also the play time significantly increases \textit{during} the competition period, as demonstrated by the ALP metrics which reaches values over 65 minute/player (cf. Table~\ref{table:gwap-metrics}).

Those results prove that providing a tangible  reward to players can make them contribute more efficiently, speeding up the classification process (higher throughput),  engaging them for a longer time (higher ALP), and ensuring a larger contribution rate to the human computation task (higher EC). As a global result, more tasks get solved.

\subsection{[Q2] Do the extrinsic reward effects last over time?}

Adding a tangible prize to a game does not seem to ensure lasting effects. In Night Knights, looking deeper in the \textit{before} and \textit{after} periods in Table~\ref{table:gwap-metrics}, we do not notice substantial differences in terms of classification and participation rate. The metrics of the \textit{before} period are slightly higher, probably due to the fact that more users tried the game, attracted by advertising campaigns (small peaks in Figure~\ref{fig:classified-per-day}) and by the novelty of the game. 

Given this similarity, in our analysis we think it worth distinguishing only between \textit{intrinsic motivation} periods (e.g., Night Knights \textit{before} and \textit{after} periods together, when users play only to have fun) and \textit{extrinsic motivation} periods (e.g., the \textit{during} phase of Night Knights, with the tangible and valuable reward). 

\subsection{[Q3] Does playing style change with the incentive?}

Defining \emph{contribution speed} the number of images played in each round, we check if also this metrics is influenced by a tangible reward. 

As explained in Section~\ref{sec:NK}, each round in Night Knights lasts one minute and each user is asked to classify one image at a time, so users have to be quick and classify as many images as possible to increase their score and being successful in the game. Given the image loading time, connection delays and waiting time for the other player's answer, we estimate that in this case classifying each image takes at least 3--5 seconds, which means 12--20 photos per round.

\begin{figure}[h]
	\hfill
	\subfigure[Extrinsic motivation]{\includegraphics[width=.49\columnwidth]{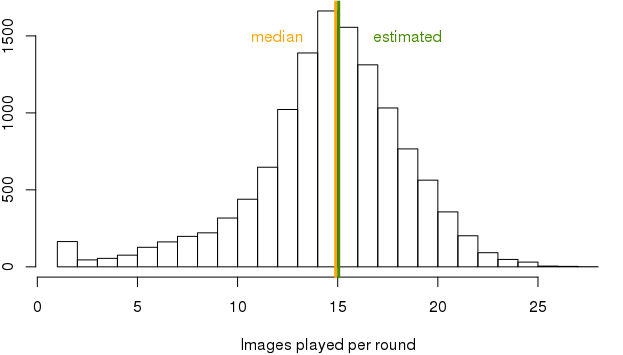}} 
	\hfill
	\subfigure[Intrinsic motivation]{\includegraphics[width=.49\columnwidth]{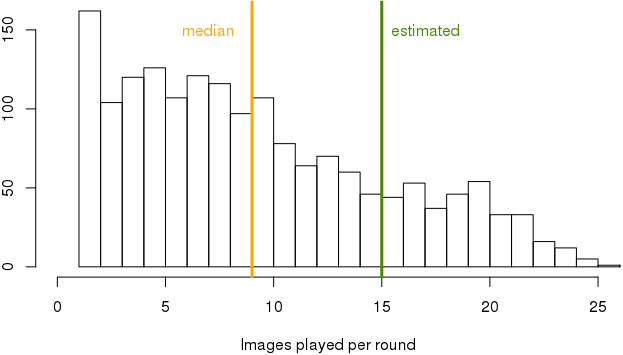}}
	\hfill
	\caption{Distribution of the number of images played in each round}
	\label{fig:img-per-round}
	\vspace{-.3cm}
\end{figure}

As Figure~\ref{fig:img-per-round} shows, in the \textit{extrinsic motivation} period, the contribution speed follows a normal distribution centered around 15 photos/round, while, in the \textit{intrinsic motivation} phase, the distribution is flat and most players played less than 10 images/round. This indicates that, during the competition, all players did their best to classify as many images as possible, reaching a median value of 15 that coincides with the estimated image classification time. On the other hand, in the \textit{intrinsic motivation} period, people play the game in a more ``relaxed'' way, just to try and explore it, taking more  time to answer. 

\section{Applying Citizen Science Engagement profiles}\label{sec:engagement}

As a first step to the assessment of player behaviour, we adopt the \emph{engagement metrics} proposed by~\cite{ponciano2015finding}: 
\textbf{activity ratio}, number of days a user plays a game divided by the total number of days the user remains linked to the game; 
\textbf{daily devoted time}, average time (e.g. in hours) a user plays the game in each active day; 
\textbf{relative active duration}, ratio of days during which a player remains linked to the game and the total number of days since the player joined the game until the day the game is over (this metric can be computed only if a ``game end'' is envisaged, which is not always the case in GWAPs); 
and \textbf{variation in periodicity}, standard deviation of the intervals between each pair of non-consecutive active days. 
Computing those metrics for each player and then applying clustering techniques leads to the identification of \emph{engagement profiles}. Our goal is to assess if the profiles recognized in citizen science literature with respect to volunteer behaviour are also detected in GWAP player behaviour and if player profiles are affected by game incentives.
Indeed, we expect player behaviour to differ from volunteer engagement. 

\subsection{[Q4] How does GWAP behaviour compare to traditional citizen science engagement?}
The mean values (and in brackets standard deviation) of the four main \emph{engagement metrics} defined by~\cite{ponciano2015finding} are shown in Table~\ref{tab:engage}. For Night Knights, we distinguish the global values and those measured during the competition only (extrinsic motivation period); for comparison, we also report the values for the citizen science initiatives illustrated in~\cite{ponciano2015finding,aristeidou2017profiles}. Daily devoted time for Night Knights is measured by approximation, multiplying the number of game rounds per 1-minute duration (the actual time is higher, because players also browse leaderboards, badges, played pictures, etc.); relative active duration is computed only during the competition time, where a ``project finish time'' is defined with the contest deadline.

\begin{table}[b]
\centering
\begin{tabular}{lccccc}
\toprule
                              & \multicolumn{2}{c}{\textbf{Night Knights}}          & \textbf{MW}                & \textbf{GZ}                & \textbf{WI}                     \\
                              & global                  & compet.                   & \cite{ponciano2015finding} & \cite{ponciano2015finding} & \cite{aristeidou2017profiles}  \\
\midrule                  
\textbf{Activity ratio}       & \textbf{0.96} (0.17)    & \textbf{0.95} (0.16)      & 0.40 (0.40)                & 0.33 (0.38)                & 0.32 (0.35)                    \\
\textbf{Daily devoted time}   & 0.68 (1.94)             & \textbf{1.80} (3.30)      & 0.44 (0.54)                & 0.32 (0.40)                & --                             \\
\textbf{Rel. active duration} & --                      & \textbf{0.54} (0.35)      & 0.20 (0.30)                & 0.23 (0.29)                & 0.43 (0.44)                    \\
\textbf{Var. in periodicity}  & 14.53 (17.9)            & \textbf{2.53} (2.12)      & 18.27 (43.3)               & 25.23 (49.2)               & 5.11 (5.36)                    \\
\bottomrule \\  
\end{tabular}
\caption{Engagement metrics (mean values and standard deviation in brackets): comparison of Night Knights (global values and competition-only metrics) with citizen science campaigns (MW: Milky Way, GZ: Galaxy Zoo, WI: Weather-it).} 
\label{tab:engage}
\end{table}

We observe that Night Knights players display quite a different behaviour with respect to volunteers: they show a 2-3 times higher activity ratio, and also consistently higher values for daily devoted time and relative active duration; this may mean that GWAP players tend to contribute in a more regular manner than volunteers. Focusing on the competition, those metrics also show a clear increase in engagement, with a significantly lower value of variation in periodicity, which suggests that the limited-time contest period stimulates players to access the game even more frequently and regularly.

Clustering players to identify engagement profiles does not give the same results as in the cited citizen science analyses~\cite{ponciano2015finding,aristeidou2017profiles}. Cross-validation between different methods (within groups sum of squares and Silhouette statistics) suggests an optimal clustering with 3 groups. Applying both agglomerative hierarchical clustering and K-means clustering yields to similar and very unbalanced grouping, with one big cluster (around 90\% of players) roughly corresponding to the \emph{hardworker} profile (high activity ratio and low variation in periodicity); the remaining players are grouped in a small cluster that we can name \emph{``focused'' hardworkers} (similar to hardworkers but with higher daily devoted time) and another small cluster that does not clearly correspond to known profiles (low values of all metrics, but higher variation in periodicity). The spasmodic, persistent, lasting and moderate profiles defined in~\cite{ponciano2015finding} are not observed. This can be interpreted as another difference between players and volunteers engagement, with game users either heavily playing and contributing, or simply trying out the game without being actually engaged.

\subsection{[Q5] What does player behaviour tell about the game nature?}

If we also evaluate user engagement in terms of when players participated, i.e. for how long they played the game, from the first to the last played round, we discover that only few users played the game both in the intrinsic motivation and extrinsic motivation periods; in particular, only 13 users played both \textit{before} and \textit{during} the competition and only 17 users became aware of the existence of the game during the competition and went on playing it \textit{after} its end. 

In addition, by analysing the users' total active time (difference between the last and the first time a user played the game), we discover that most of the users played for a very short amount of time; 75\% of players used the game for less than 5 minutes and only the 10\% played for more than a day. 

These statistics are not surprising, because they are strong indicators of the game nature, which is a so-called \emph{casual game}. 
Casual games are usually designed to be played in short bursts of a few minutes and then set aside. By their very nature, casual games target the short free/leisure time  between the myriad of  everyday tasks, such as between work and domestic obligations or between attention and distraction~\cite{anable2013casual}.  
Regarding the overall time spent playing mobile games, the literature shows that an average gamer spends every day approximately 24 minutes playing games on mobile devices, with \textit{heavy gamers} spending about 1 hour/day and \textit{light gamers} about 2 minutes/day~\cite{hwong2016leveling}. 
 
\section{Defining GWAP Engagement profiles}\label{sec:analysis2}

Given that volunteer profiles in citizen science do not seem to suitably describe GWAP players, we focus our investigation on two additional main metrics, player accuracy and player participation, more closely related to human computation, and analyse their interplay with different factors, like game incentive, task difficulty and task variety. The goal is to uncover GWAP-specific user behaviours and to identify \emph{GWAP-specific player profiles}.

\textbf{Player accuracy} is measured ex-post by counting how many tasks each user correctly solved over the total number of tasks he/she played with; in this context, ``correct'' refers to the final task solution computed by the truth inference algorithm. Accuracy takes values between 0 and 1 and corresponds to the worker precision or labeling quality metrics used in crowdsourcing literature (e.g.~\cite{zheng2017truth}).
\textbf{Player participation} is measured as the total number of contributions given by each user in  the game rounds he/she played. While there are of course alternative ways to measure participation (e.g., number of game rounds, total played time), we prefer to consider the number of contributions, since this indicator is more closely related to the ``task'' execution and the game purpose.

\subsection{[Q6] What kind of GWAP player profiles can be identified?}
\begin{figure}[b!]
\centering
	\includegraphics[width=.7\columnwidth]{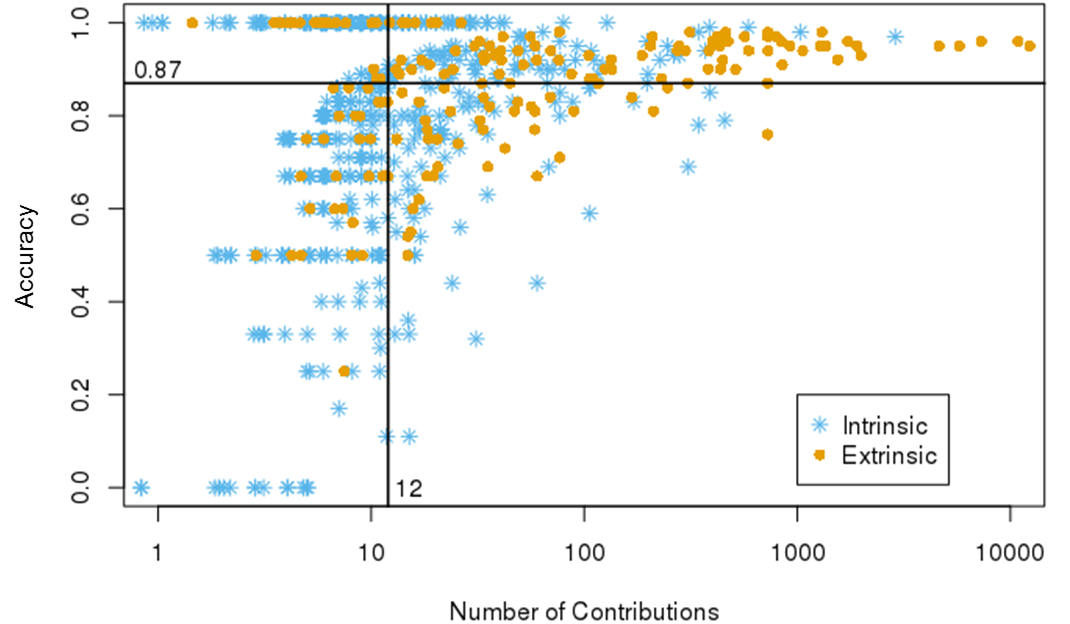}
	\caption{Players' participation vs. accuracy and median values}
	\label{fig:4quadrants}
\end{figure}

Referring again to Night Knights data, we plot each user as a data point along participation and accuracy axes (cf. Figure~\ref{fig:4quadrants}). 
To divide players into groups, we applied clustering as in Section~\ref{sec:engagement}, but -- at least in the case of Night Knights -- the results put 98-99\% of players in the same cluster, placing only ``outliers'' in the other clusters. Therefore, to define GWAP-specific profiles, we propose to simply set separation thresholds on the two axes dividing the space into quadrants; more specifically, we adopt the \emph{median} as separation value, which is a commonly used measure and robust statistic. While this definition is arbitrary, it is also data-independent, thus the proposed approach can be adopted to analyse and compare different GWAPs without loss of generality.

The thresholds calculated on the Night Knights dataset are 12 contributions for the x-axis and 0.87 accuracy for the y-axis. The median value for participation roughly corresponds to the separation between those who played just a couple of game rounds from those who were more deeply engaged (cf. Section~\ref{sec:analysis1}). The median accuracy value is quite high and this is a good sign about the GWAP efficacy to achieve its purpose; in other cases, when a specific minimum value of accuracy is required, the threshold choice could be driven by domain-specific consideration instead of being identified by the median.

By using this approach, the investigation space is divided into areas that represent different ``behavioral'' profiles as follows. 
\begin{figure}[t]
	\centering
	\includegraphics[width=.9\columnwidth]{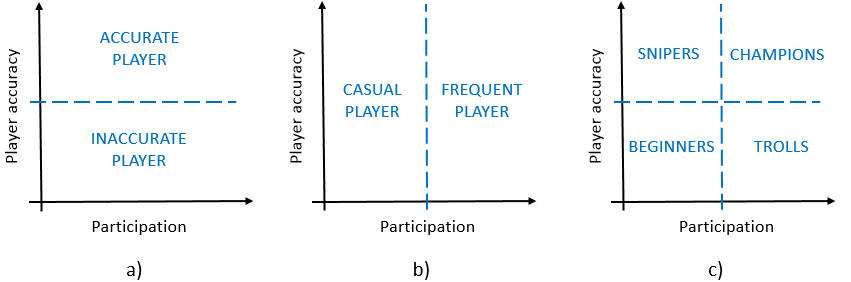}
	\caption{Definition of GWAP-specific player profiles}
	\label{fig:user-profiles}
\end{figure}
Along the accuracy axis, we obtain two profiles: \textit{accurate players}, i.e. players with an accuracy higher than the median, and the remaining \textit{inaccurate players} (cf. Figure~\ref{fig:user-profiles}-a). 
Along the participation axis (cf. Figure~\ref{fig:user-profiles}-b), we define \textit{casual players} those who contribute less than the median, and \textit{frequent players} the most addicted and loyal contributors. 
Considering both dimensions, we define four profiles (cf. Figure~\ref{fig:user-profiles}-c):
\begin{itemize}
	\item \textit{Beginners} (bottom-left): this is the set of users that play the game for a short period of time, just for curiosity; this kind of players gives only few contributions with low accuracy.   
	\item \textit{Snipers} (top-left): users that are very accurate in their contributions but they contribute only a little. Ideally, they should be motivated to become champions, since their contributions are valuable. 
	\item \textit{Champions} (top-right): this is the most desirable category of players, since they have high level of participation with very high accuracy. 
	\item \textit{Trolls} (bottom-right): this is the category of less desirable users, since they give a lot of inaccurate contributions; having a lot of \textit{Trolls} in the game either makes the classification process longer, since it is harder to reach an agreement, or even leads to undesired results.
\end{itemize}

\begin{figure}[b]
		\centering
  	\includegraphics[width=.7\columnwidth]{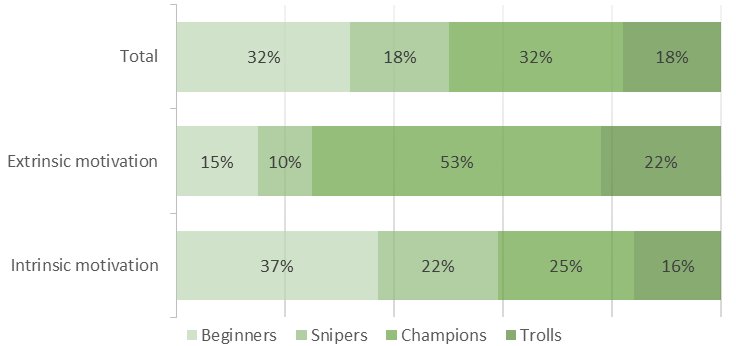}
  	\caption{Distribution of players between profiles, in total and with different incentives}
  	\label{fig:perc-quadrants-matrix}
  \end{figure}

Observing again Night Knights data, we can also quantitatively analyse the effect of game incentive on the profile composition (cf. Figure~\ref{fig:perc-quadrants-matrix}). 
With extrinsic motivation, most users (53\%) acted as {champions}, and this share is much higher than in the total (32\%). 
On the other hand, with the intrinsic motivation only, the presence of {champions} was lower, only 25\%. This difference may indicate that the different incentives lead to different user behaviour; the presence of tangible rewards can engage users for a longer time and can motivates them to contribute with more effort and attention.  

With intrinsic motivation, also the percentages of {snipers} was higher than the average. 
The largest group of users in the intrinsic motivation period, however, was {beginners} (37\%): probably this happened because they tried the game just for curiosity or to understand how the game works, without paying too much attention to the answers they gave. 
As expected, the number of {beginners} was very low with the extrinsic motivation, since they had a clear goal to play the game.  
Fortunately, the percentages of {trolls} were low in both periods. This means that the Night Knights game succeeded in avoiding too many spammers that could have made the classification process longer or more inaccurate.

While the above results are specific to Night Knights, the profile analysis can be applied to any other GWAP; indeed, examining the composition of a GWAP player population can reveal different behaviour and inform game re-design.
 
Finally, we would like to point out an insight that is not immediately evident in Figure~\ref{fig:4quadrants}: since the players on the right part of the plot are those who contributed more, if we sum the contributions from the four profiles, we obtain the figures in Table~\ref{tab:perc-contr-row}. In the case of our GWAP, therefore, the large majority of contributions comes from the most active and accurate players, which is reassuring with respect to the achievement of the game purpose. 

\begin{table}[h]
	\centering
	\begin{tabular}{lcccc}
		\toprule
		                   & ~\textbf{Beginners} & ~\textbf{Snipers} & ~\textbf{Champions} & ~\textbf{Trolls} \\
		Task contributions &  0.7\%              &  0.4\%            &  95.9\%             &  3.0\%           \\ 
		\bottomrule \\
	\end{tabular}
	\caption{Distribution of contributions across players profiles}
	\label{tab:perc-contr-row}
	\vspace{-.5cm}
\end{table}

In the following, we analyse the interplay between player accuracy and player participation by taking into account additional factors. More specifically, we check if there is a statistically significant difference between the mean accuracy of \textit{casual} and \textit{frequent players} with respect to some control variables, namely the incentive type, the task difficulty and the task variety. 

\subsection{[Q7] Does  player behaviour change with different incentives?}
To answer this question, we check for mean difference in accuracy for casual and frequent players in the intrinsic and extrinsic motivation periods.

\begin{figure}[t]
	\centering
	\hfill
	\subfigure[Extrinsic]{\includegraphics[width=.24\columnwidth]{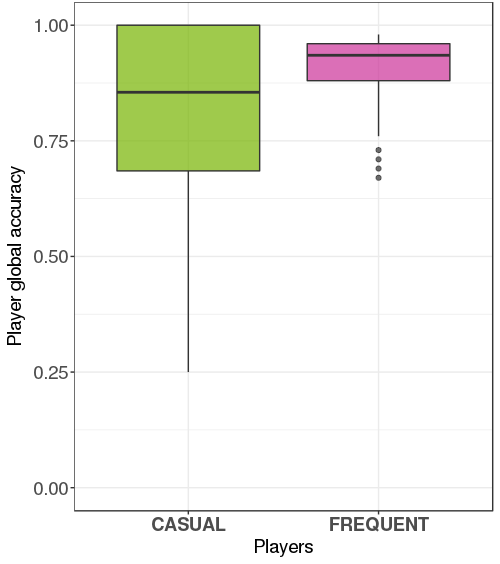}}
	\hfill
	\subfigure[Intrinsic]{\includegraphics[width=.24\columnwidth]{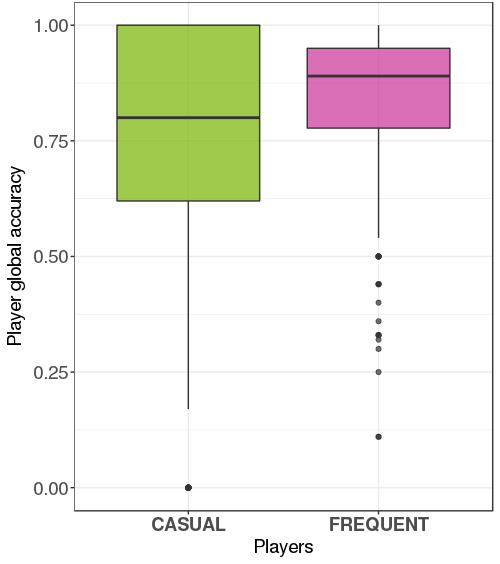}}
	\hfill
	\subfigure[Easy]{\includegraphics[width=.24\columnwidth]{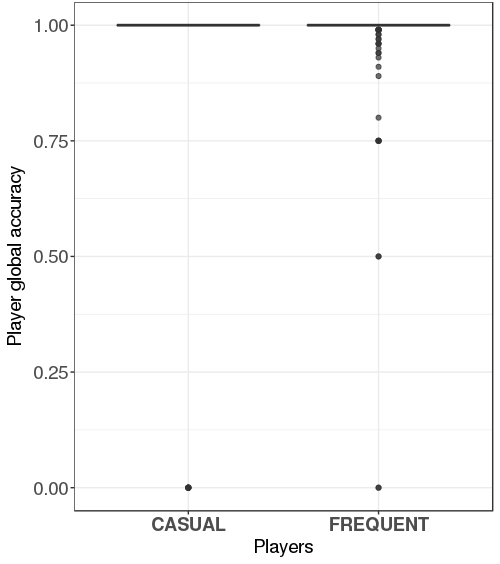}}
	\hfill
	\subfigure[Difficult]{\includegraphics[width=.24\columnwidth]{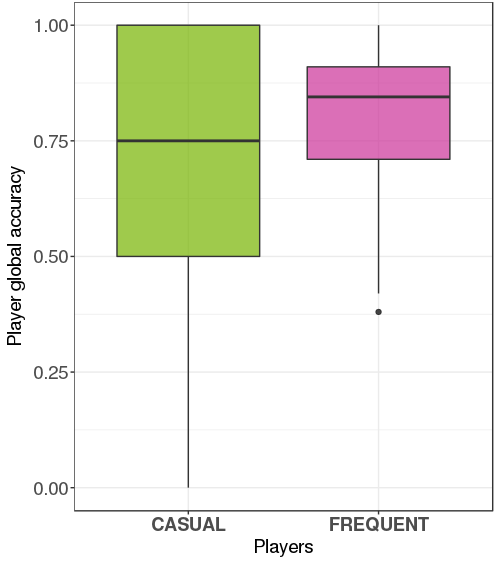}}
	\hfill
	\caption{Accuracy distribution of \textit{casual} and \textit{frequent} players with different incentives (\emph{a} and \emph{b}) and with different task difficulty (\emph{c} and \emph{d}). The difference between players' profiles is statistically significant in all cases except for easy tasks. }
	\label{fig:casual-frequent}
\end{figure}

In Night Knights, the average accuracy of the \textit{frequent players} is higher than the one of \textit{casual players} in both  periods, as shown in the first two boxplots of Figure~\ref{fig:casual-frequent}; this difference is also significant from a statistically point of view (p-value of the t-test less than 0.05). 
We also notice a mean accuracy increase of about 10\% when a tangible rewards is present (from 0.74 to 0.81 for casual and from 0.83 to 0.90 for frequent): since during the competition users were  encouraged to play to win the prize, they  paid more attention to the image classification, raising also the answers' quality. 

This may indicate that in GWAPs \textit{frequent players} contribute in a more accurate way than \textit{casual} ones, and that extrinsic motivation has a positive impact on accuracy. 

\subsection{[Q8] Does  player behaviour change with task difficulty?}
We define \emph{task difficulty} as the number of different users  needed to solve it (the higher the number, the harder the task); this is because our incremental truth inference algorithm (cf. Sec\-tion~\ref{sec:NK}) dynamically estimates the number of contributions required to solve a task. 
We split the images in two sets based on their difficulty and we check if this impacts player behaviour.

For Night Knights, we marked as ``easy'' the images that requires only 4 contributions (the minimum number to reach an agreement according to our domain experts), and as ``difficult'' those that required more contributions. 
``Easy'' images are 58\% of all classified images, while the number of contributions required to classify ``difficult'' images ranges from 5 to 17. 

As shown in the (c) and (d) boxplots in Figure~\ref{fig:casual-frequent}, accuracy on ``easy'' images is almost the same  between \textit{casual} and \textit{frequent players} (indeed, the difference in mean accuracies is not statistically significant). 
On the contrary, this difference is statistically significant for ``difficult'' images (mean accuracy is 0.84 for \textit{frequent players} and 0.68 for \textit{casual players}). 

Those results suggest a \emph{learning effect} in GWAPs: the more a user plays the game, the more he/she understands the task to be solved, thus increasing his/her accuracy and consequently also result quality. 

\subsection{[Q9] Does  player behaviour change with task variety?}
Since Night Knights aims to solve a multinomial classification task, we investigate whether there is any evident phenomenon related to the different image categories. Therefore, we compute again the accuracies of the two groups of casual and frequent players in classifying the 6 output classes. 
We summarize the mean accuracy values in Table~\ref{tab:variety}.  
 
\begin{table}[t]
	\centering
	\footnotesize
	\begin{tabular}{@{}lcccccc@{}}
		\toprule
		& \multicolumn{1}{l}{\textbf{Black}} & \multicolumn{1}{l}{\textbf{City}} & \multicolumn{1}{l}{\textbf{Stars}} & \multicolumn{1}{l}{\textbf{Aurora}} & \multicolumn{1}{l}{\textbf{ISS}} & \multicolumn{1}{l}{\textbf{None}} \\ \midrule
		\textbf{Casual}   & 0.69 & 0.88 & 0.57 & 0.74 & 0.63 & 0.70                              \\
		\textbf{Frequent} & 0.79 & 0.91 & 0.68 & 0.77 & 0.77 & 0.77                              \\ 
		\bottomrule \\
	\end{tabular}
	\caption{Mean accuracy of \textit{casual} and \textit{frequent} players with images of different categories. The difference is not statistically significant for any of the categories. }
	\label{tab:variety}
	\vspace{-.5cm}
\end{table}

Applying the t-test to check if the mean accuracy is different for the two players' profiles, we cannot reject the null hypothesis. This may mean that any player is equally able/unable to distinguish the different categories, independently of his/her level of participation; indeed, in our GWAP, there is no need for background- or domain-specific knowledge to play the game. This analysis can help in identifying the need for training or expert knowledge of GWAP players.

\begin{figure}[h]
	\centering
	\includegraphics[width=.65\columnwidth]{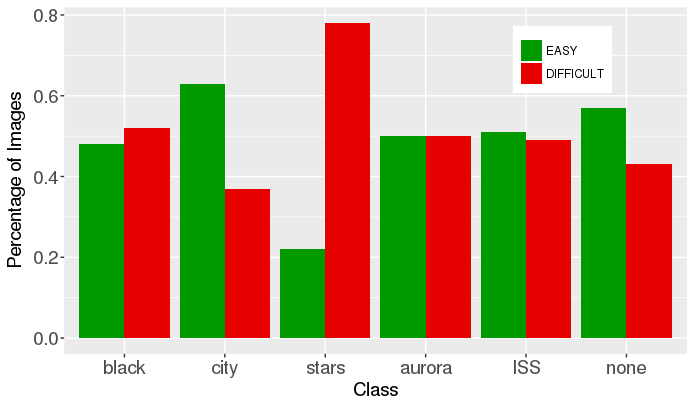}
	\caption{Distribution of easy/difficult tasks across different image categories.}
	\label{fig:barplot-easy-diff}
	\vspace{-.5cm}
\end{figure}

On the other hand, the mean accuracy values change a lot across different categories, spanning between 0.57 and 0.91. This is also explained by the different distribution of easy/difficult tasks across the variety of classes, as shown in Figure~\ref{fig:barplot-easy-diff}. Indeed, some categories are intrinsically more difficult to classify than others, but Table~\ref{tab:variety} shows that this complexity related to task variety is equally perceived by players with low and high levels of participation.

\section{Conclusions}\label{sec:concl}

In this paper, we presented an investigation of the interplay of different factors in the evaluation of GWAP results. More specifically, we focused on the profiling of players according to different user metrics 
and we studied the influence of game incentive and task characteristics. 

To inform our discussion, we described the results of such multi-dimensional analysis over the data collected by a GWAP for multinomial classification of images. 
While some of our considerations result from the quantitative analysis of a single game, and are not \emph{per se} generalizable, we believe that the proposed approach is replicable to evaluate any other GWAP. We believe that such deeper analysis is an important (and sometimes neglected) investigation to understand players' behaviour, to evaluate the impact of various factors on reliability and quality, and finally to assess the ability of GWAPs to achieve their intended purpose and its sustainability over time.

Finally, we would like to point out that, even when player participation is limited in time, a classification GWAP can be used 
to build a reasonably large training set to be used in traditional machine learning settings to train classifiers for  larger-scale labeling. 
In our previous work, we showed that humans and machines indeed agree on image classification for the Night Knights dataset~\cite{re2018human}.

\vspace{-.25cm}
\subsection*{Acknowledgments}
\small
This work is partially supported by the STARS4ALL project (H2020-688135), co-funded by the European Commission. We thank all the Night Knights players who contributed to the classification task solution and allowed us to perform this work.
\vspace{-.25cm}

\bibliographystyle{splncs}
\bibliography{biblio} 

\end{document}